\documentclass[pra,twocolumn,showpacs]{revtex4-1}
\usepackage{amsmath,amscd,amssymb,color}
\usepackage{graphicx,amsfonts,dsfont}
\usepackage{epstopdf}
\usepackage{hyperref}
\usepackage{enumerate,bbold}

\begin{document}
\title{Experimental demonstration of fully contextual quantum 
correlations on an NMR quantum information processor}
\author{Dileep Singh}
\email{dileepsingh@iisermohali.ac.in}
\affiliation{Department of Physical Sciences, Indian
Institute of Science Education \& 
Research Mohali, Sector 81 SAS Nagar, 
Manauli PO 140306 Punjab India.}
\author{Jaskaran Singh}
\email{jaskaransinghnirankari@iisermohali.ac.in}
\affiliation{Department of Physical Sciences, Indian
Institute of Science Education \& 
Research Mohali, Sector 81 SAS Nagar, 
Manauli PO 140306 Punjab India.}
\author{Kavita Dorai}
\email{kavita@iisermohali.ac.in}
\affiliation{Department of Physical Sciences, Indian
Institute of Science Education \& 
Research Mohali, Sector 81 SAS Nagar, 
Manauli PO 140306 Punjab India.}
\author{Arvind}
\email{arvind@iisermohali.ac.in}
\affiliation{Department of Physical Sciences, Indian
Institute of Science Education \& 
Research Mohali, Sector 81 SAS Nagar, 
Manauli PO 140306 Punjab India.}
\begin{abstract}
The existence of contextuality in quantum  mechanics is a fundamental departure
from the classical description of the world.  Currently, the quest to identify
scenarios which cannot be more contextual than quantum theory is at the
forefront of research in quantum contextuality.  In this work, we
experimentally test two inequalities, which are capable of revealing fully
contextual quantum correlations,  on a Hilbert space of dimension eight 
and four
respectively, on an NMR quantum information processor.  The projectors
associated with the contextuality inequalities are first reformulated in terms
of Pauli operators, which can be determined in an NMR experiment.  We also
analyze the behavior of each inequality under rotation of the underlying
quantum state, which unitarily transforms it to another pure state.
\end{abstract} 
\maketitle 
\section{Introduction}
\label{sec:intro}
Non-contextual hidden variable (NCHV) theories in which outcomes of measurements
do not depend on other compatible measurements, have been shown not to reproduce
quantum correlations~\cite{bell-rmp-66,specker-jmm-67}.  Quantum mechanics (QM)
exhibits the property of
contextuality~\cite{jaskaran-pra-17,raussendorf-pra-13,howard-nature-14} which
implies that measurement results of observables depend upon other commuting
observables which are within the same measurement test. Much recent research is
going on in the direction of guessing the physical principle responsible for
this form of contextuality~\cite{cabello-nature-11}.  The pertinent questions
that arise include whether there is any theory more contextual than quantum
mechanics and whether the simplest scenario in which more general theories
cannot be more contextual than quantum mechanics can be
identified~\cite{cabello-pra-13,chiribella-pra-11,barnum-prl-10,pawlowski-nature-09}.

Contextuality tests correspond to the violation of certain inequalities
involving expectation values, and the first such test was proposed by Kochen and
Specker~\cite{specker-jmm-67} by using a single qutrit system (the KS theorem),
and a modified KS scheme was constructed by Peres~\cite{peres-jpamg-91}.
State-independent~\cite{plastino-pra-10,badzia-prl-09,cabello-prl-15} tests use
the set of observables such that for any quantum state there is no probability
distribution which can describe the outcome of measurement of these observables
on that state, hence these tests are able to reveal the contextual behavior of
any state of the quantum system. On the other hand, the
state-dependent~\cite{klyachko-prl-08,kurzynski-pra-12,sohbi-pra-16} tests
typically use fewer observables to show that no joint probability distribution
can describe the measurement outcomes on a certain subset of states of the
quantum system. The smallest indivisible physical system exhibiting quantum
contextuality for repeatable measurements is a qutrit (a three-level quantum
system)~\cite{bell-rmp-66}. The simplest state-dependent  non-contextual
inequality which is commonly referred to as the
Klyachko-Can-Binicioglu-Shumovsky (KCBS) inequality~\cite{klyachko-prl-08}, for
a qutrit requires five experiments, each of them involving two compatible yes-no
tests~\cite{cabello-pra-13}.  Several experimental tests of quantum
contextuality have been demonstrated by different groups using
photons~\cite{zu-prl-12,ambrosio-prx-13,amselem-prl-09,nagali-prl-12,huang-pra-13},
ions~\cite{zhang-prl-13,leupold-prl-18}, neutrons~\cite{bartosik-prl-09} and
nuclear spins~\cite{moussa-prl-10,dogra-pla-16}.

In this paper, we experimentally demonstrate fully contextual quantum
correlations via two different inequalities, on an NMR quantum information
processor. The first inequality as proposed by Cabello~\cite{cabello-pra-13},
utilizes ten projectors and requires five measurements on a state in a Hilbert
space of dimension at least six. We demonstrate this inequality by realizing the
six-dimensional subspace on states in an eight-dimensional Hilbert space.  The
second inequality as proposed by Nagali {\em et.~al}~\cite{nagali-prl-12}, uses
ten projectors and ten measurements which we implement on states in a
four-dimensional Hilbert space. For experimental verification of both the
inequalities, we decompose all the projectors involved in terms of Pauli
operators.  The advantage is two-fold: first, it reduces the need of performing
quantum state tomography which is a resource-intensive procedure and second, the
inequalities can be tested by using a fewer number of observables.  The
eight-dimensional and four-dimensional Hilbert spaces are physically realized
using three and two NMR qubits, respectively.  Violation of the inequalities as
observed experimentally match well with theoretical predictions and have an
experimental fidelity  $\geq 0.96$. We also study the behavior of both the
inequalities when the underlying quantum state undergoes a rotation.  Our
results imply that the violation of both inequalities follows a nonlinear trend
with respect to the rotation angle of the underlying state. We also find that
fully contextual quantum correlations on an eight-dimensional Hilbert space are
more robust against state rotation, as compared to the ones on the
four-dimensional Hilbert space, allowing a greater angle for violation.

The material in this paper is arranged as follows: Section~\ref{ineq1}
describes the fully contextual quantum correlations, 
the quantum state and the yes/no
tests required to reveal correlations with zero non-contextual content and their
experimental implementation on an eight-dimensional quantum system using three
NMR qubits.  Section~\ref{ineq2} describes fully contextual quantum
correlations  
in a four-dimensional Hilbert space,
and its experimental implementation 
using two NMR qubits.  Section~\ref{concl} contains a few
concluding remarks. 
\section{Fully contextual quantum correlations in 
an eight-dimensional Hilbert space}  
\label{ineq1}
In this section, we first review a contextuality inequality which is capable of
revealing fully contextual quantum correlations as developed by
Cabello~\cite{cabello-pra-13}, which requires a Hilbert space dimensionality of at least
$6$.  We then design a modified version of the inequality via decomposition
of the projectors 
into Pauli matrices, for ease of experimental implementation.  We
experimentally test the inequality on an eight-level quantum system,  physically
realized via three NMR qubits.
\begin{figure}[ht]
\centering
\includegraphics[scale=1.0]{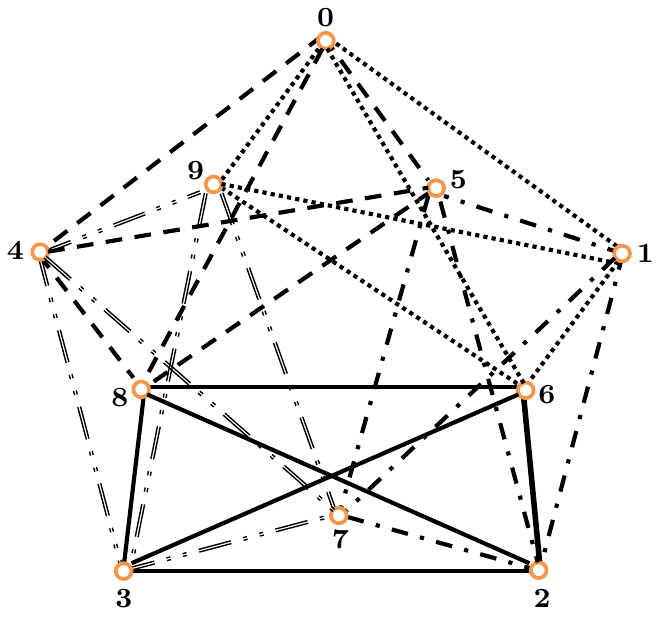}
\caption{Orthogonality graph corresponding to the KCBS-twin inequality $\mathcal{K}$. Vertices correspond to projectors, while edges represent
the orthogonality relationship between two vertices. 
Five sets of four interconnected vertices correspond to 
measurements involved in testing $\mathcal{K}$ and are differentiated 
by different edge line styles.}
\label{fig-twin-kcbs}
\end{figure}
\begin{table}[h]
\caption{Pauli operators for 
three qubits, used to decompose the
corresponding projectors for the experimental demonstration of the inequality $\mathcal{K}$.}
\vspace*{12pt}
\centering
\renewcommand{\arraystretch}{1.7}
\begin{tabular}{cc}
\hline
  Pauli operators  & 
Pauli operators ~~~\\
\hline
 $A_{0}$ = $ \mathds{1} \otimes \mathds{1} \otimes \sigma_x$ &  $A_{18}$ = $ \sigma_x  \otimes\sigma_z\otimes \sigma_x$~~~\\
$A_{1}$ =$ \mathds{1} \otimes \mathds{1} \otimes \sigma_z$ & $A_{19}$ = $ \sigma_x  \otimes\sigma_z\otimes \sigma_z$~~~\\
 $A_{2}$ = $\mathds{1} \otimes  \sigma_x \otimes \mathds{1}$ &  $A_{20}$ = $ \sigma_y  \otimes \mathds{1} \otimes \sigma_y$ ~~~\\
$A_{3}$ = $\mathds{1} \otimes  \sigma_x \otimes \sigma_x$ & $A_{21}$ = $ \sigma_y  \otimes\sigma_x\otimes \sigma_y$ ~~~\\
$A_{4}$ = $\mathds{1} \otimes  \sigma_x \otimes \sigma_z$& $A_{22}$ = $ \sigma_y  \otimes\sigma_y\otimes \mathds{1}$~~~\\
$A_{5}$ = $\mathds{1} \otimes  \sigma_y \otimes \sigma_y$ & $A_{23}$ = $ \sigma_y  \otimes\sigma_y\otimes \sigma_x$ ~~~\\
$A_{6}$ = $\mathds{1} \otimes  \sigma_z \otimes \mathds{1}$ &$A_{24}$ = $ \sigma_y  \otimes\sigma_y\otimes \sigma_z$ ~~~\\
 $A_{7}$ = $\mathds{1} \otimes  \sigma_z \otimes \sigma_x$  &  $A_{25}$ = $ \sigma_y  \otimes\sigma_z\otimes \sigma_y$  ~~~\\
$A_{8}$ = $\mathds{1} \otimes  \sigma_z \otimes \sigma_z$ & $A_{26}$ = $ \sigma_z  \otimes \mathds{1} \otimes \mathds{1}$ ~~~\\
 $A_{9}$ = $ \sigma_x \otimes \mathds{1} \otimes \mathds{1}$ & $A_{27}$ = $ \sigma_z  \otimes \mathds{1} \otimes \sigma_x$ ~~~\\
$A_{10}$ = $ \sigma_x \otimes \mathds{1} \otimes\sigma_x$ & $A_{28}$ = $ \sigma_z \otimes \mathds{1} \otimes\sigma_z$ ~~~\\
$A_{11}$ = $ \sigma_x \otimes \mathds{1} \otimes\sigma_z$ &$A_{29}$ = $ \sigma_z  \otimes\sigma_x \otimes \mathds{1}$ ~~~\\
$A_{12}$ =  $ \sigma_x  \otimes\sigma_x\otimes \mathds{1}$ & $A_{30}$ = $ \sigma_z  \otimes\sigma_x \otimes \sigma_x$ ~~~\\
$A_{13}$ = $ \sigma_x  \otimes\sigma_x\otimes \sigma_x$ & $A_{31}$ =  $ \sigma_z  \otimes\sigma_x \otimes \sigma_z$  ~~~\\
$A_{14}$ = $ \sigma_x  \otimes\sigma_x\otimes \sigma_z$ &  $A_{32}$ = $ \sigma_z  \otimes\sigma_y \otimes \sigma_y$ ~~~\\
 $A_{15}$ = $ \sigma_x  \otimes\sigma_y\otimes \sigma_x$ & $A_{33}$ =  $ \sigma_z  \otimes\sigma_z \otimes \mathds{1} $  ~~~\\
$A_{16}$ = $ \sigma_x  \otimes\sigma_y\otimes \sigma_y$ & $A_{34}$ = $ \sigma_z  \otimes\sigma_z \otimes \sigma_x$~~~\\
$A_{17}$ = $ \sigma_x  \otimes\sigma_z\otimes \mathds{1}$& $A_{35}$ =  $ \sigma_z  \otimes\sigma_z \otimes \sigma_z$ ~~~\\
\hline
\end{tabular}
\label{tab:ais}
\end{table}

The simplest test of quantum contextuality requires 
the measurement of five
different projectors $\lbrace\Pi_i\rbrace$, $i\in\lbrace 0,1,2,3,4\rbrace$ and
$\Pi_i=|v_i\rangle\langle v_i|$, where $|v_i\rangle$ are 
unit vectors~\cite{klyachko-prl-08}. These
projectors follow the exclusivity relation
$P(\Pi_i=1)+P(\Pi_{i\oplus 1}=1) = 1$, where $P(\Pi_i=1)$ represents the
probability of obtaining the outcome $\Pi_i$, 
and addition is taken modulo five. For
projective measurements, this relationship implies that only one of $\Pi_i$ or
$\Pi_{i\oplus 1}$ can be obtained in a joint measurement of both. The
corresponding test, termed as KCBS inequality~\cite{cabello-pra-13}
is of the form
\begin{equation}
\frac{1}{2}\sum_{i=0}^{4} P(\Pi_i+\Pi_{i\oplus 1}=1)\overset{\text{NCHV}}{\leq} 2 \overset{\text{QM}}{\leq} \sqrt{5} \overset{\text{GP}}{\leq} \frac{5}{2},
\label{eq:kcbs}
\end{equation}
where the inequalities correspond to the maximum value achievable for
non-contextual hidden variable (NCHV) theories, quantum mechanics (QM) and
generalized probabilistic (GP) theories.
\begin{table}[h]
\label{3qubit-Pauli-table}
\caption{Product operators for a three-qubit
system, mapped to the Pauli $z$ operators via the
initial state transformation $\rho \rightarrow \rho_i=U_i.\rho.U^\dagger_i$.}
\vspace*{12pt}
\centering
\renewcommand{\arraystretch}{1.7}
\begin{tabular}{rp{12pt}l}
\hline
 Observable Expectation  & &
Unitary Operator~~~\\
\hline
 $\langle \sigma_{3z} \rangle$ = Tr[$\rho_{1}.\sigma_{3z}$]
&&  $U_{1}$=Identity ~~~\\
 $\langle \sigma_{2z} \rangle$ = Tr[$\rho_{2}.\sigma_{2z}$]
&&  $U_{2}$=Identity ~~~\\
 $\langle \sigma_{2z} \sigma_{3z} \rangle$ =
Tr[$\rho_{3}.\sigma_{3z}$] && $U_{3}=\rm{CNOT_{23}~  }$ ~~~\\
 $\langle \sigma_{1z} \rangle$ = Tr[$\rho_{4}.\sigma_{1z}$]
&&$U_{4}$=Identity ~~~\\
$\langle \sigma_{1z} \sigma_{3z} \rangle$ =
Tr[$\rho_{5}.\sigma_{3z}$]&& $U_{5}=\rm{CNOT_{13}~  }$ ~~~\\
 $\langle \sigma_{1z} \sigma_{2z} \rangle$ =
Tr[$\rho_{6}.\sigma_{2z}$] && $U_{6}=\rm{CNOT_{12}~  }$ ~~~\\
$\langle \sigma_{1z} \sigma_{2z} \sigma_{3z} \rangle$ =
Tr[$\rho_{7}.\sigma_{3z}$] && $U_{7}=\rm{CNOT_{23}~  }.\rm{CNOT_{12}~  }$ ~~~\\
\hline
\end{tabular}
\end{table}

As is evident from Eqn.~(\ref{eq:kcbs}), the maximum violation that can be achieved in quantum mechanics is less than what can be 
attained if an underlying GP model is considered. Therefore, for the KCBS scenario, quantum correlations are not fully contextual. Recently, it has 
been shown that there exist tests of contextuality for which quantum correlations saturate
the bound as imposed by 
GP models~\cite{amselem-prl-12}. For these scenarios, 
quantum correlations are either non-contextual or 
fully contextual. The simplest 
test of contextuality, capable of revealing fully contextual 
quantum correlations again requires only five measurements, 
but of ten different 
projectors $\lbrace \Pi_i\rbrace$ and is of the form,
\begin{equation}
\mathcal{K}=\frac{1}{2} \sum_{i=0}^{4} P(\Pi_i +\Pi_{i+1} +\Pi_{i+5} +\Pi_{i+7}=1)\overset{\text{NCHV}}{\leq}2\overset{\text{QM, GP}}{\leq}\frac{5}{2},
\label{eq:kcbs_twin}
\end{equation}
where the sum in the indices is defined such that $4 + 1= 0$ 
and $3 + 7= 5$. Since
both the KCBS and the aforementioned inequality (Eqn.~\ref{eq:kcbs_twin})
require only five different measurements, the above scenario is termed as a twin
inequality of KCBS, with the only difference that it is capable of revealing
fully contextual quantum correlations and requires quantum systems having
Hilbert space dimension at least six.  We will henceforth refer to this
inequality as the ``KCBS-twin'' inequality.
\begin{figure}[ht]
\centering
\includegraphics[scale=1.0]{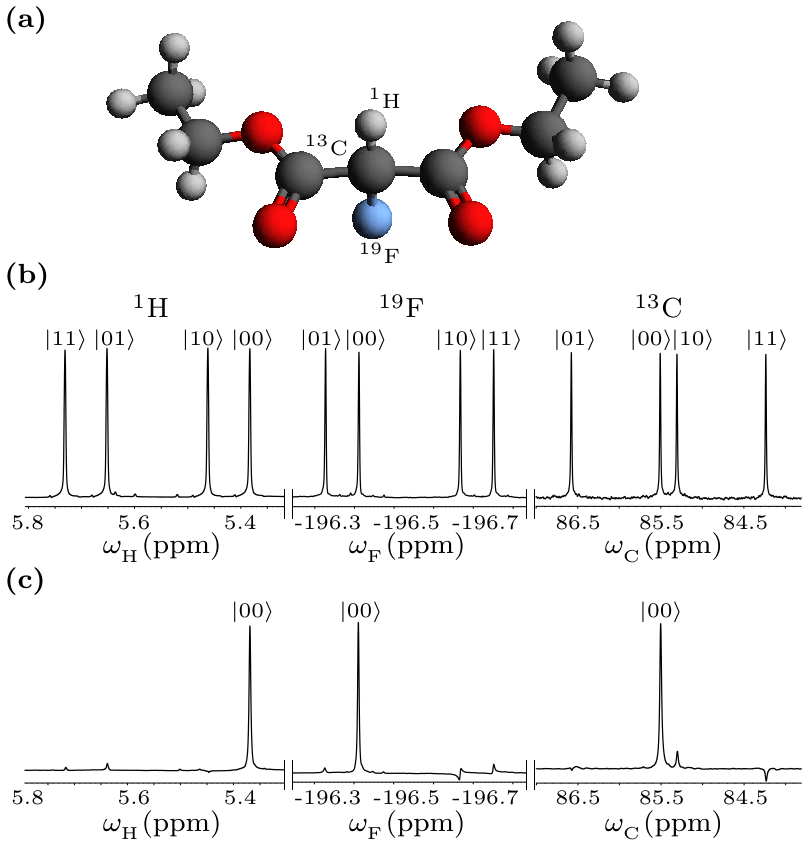}
\caption{
(a) Molecular structure of  ${}^{13}$C-labeled diethyl
fluoromalonate used to physically realize three qubits. 
NMR spectra of (b) the thermal
equilibrium state and (c) the pseudopure 
state $\vert 000\rangle$. Each peak is labeled with the
logical state of the qubit which is passive during the transition.
The horizontal
scale represents the chemical shifts in ppm.}
\label{molfig-3qubit}
\end{figure}
\begin{table}[h]
\caption{NMR parameters for 
the three-qubit ${}^{13}$C-labeled
diethyl fluoromalonate system.}
\vspace*{12pt}
\centering
\renewcommand{\arraystretch}{1.7}
\begin{tabular}{rrrrr}
\hline
Qubit & 
$\nu$ (Hz)& $J$ (Hz) & $~T_1$ (sec) & $T_2$ (sec)~~~\\
\hline
$\rm ^{1}H$ & $3334.24$ & $J_{HF}=47.5$ & $3.4$ & $1.6$~~~\\
$\rm ^{19}F$ & $~-110999.94$ & $J_{HC}=161.6$ & $3.7$ & $1.5$~~~\\
$\rm ^{13}C$ & $12889.53$ & $~J_{FC}=-191.5$ & $3.6$ & $1.3$~~~\\
\hline
\end{tabular}
\label{parameter-table-3qubit}
\end{table}

The scenario corresponding to the KCBS-twin  
inequality~(Eqn.~\ref{eq:kcbs_twin}) can be
represented by an exclusivity graph as shown in Fig.~\ref{fig-twin-kcbs}.  In
this graph, each vertex corresponds to a unit vector $|v_i\rangle$ used to
construct the projectors $\Pi_i$, and two vertices are connected by an edge if
and only if they are exclusive. From the graph it is possible to identify five
different measurements $\mathcal{M}_i$ which are defined as
\begin{equation}
\mathcal{M}_i = \lbrace \Pi_i, \Pi_{i+1}, \Pi_{i+5}, \Pi_{i+7} \rbrace, \quad \forall i\in \lbrace 0,1,...,9\rbrace.
\label{eq:twin_meas}
\end{equation}
These measurements can be identified from the graph in 
Fig.~\ref{fig-twin-kcbs} by five sets of  four
interconnected vertices, each represented by 
a different line style.

An explicit form of 
the KCBS-twin inequality~(Eqn.~\ref{eq:kcbs_twin}) 
which saturates the QM and GP bound can be obtained if we consider the unit vectors $|v_i\rangle$
defined as:
\begin{subequations}
\begin{align}
\langle v_0| \equiv &\frac{1}{\sqrt{8}} (\sqrt{2}, -\sqrt{2}, 0,0,2,0,0,0), \\
\langle v_1| \equiv &\frac{1}{\sqrt{8}} (\sqrt{2},0,0, \sqrt{2},-1,\sqrt{3},0,0), \\
\langle v_2| \equiv &\frac{1}{2}(1, -1, -1,-1,0,0,0,0),\\
\langle v_3| \equiv &\frac{1}{2} (1, -1, 1,1,0,0,0,0),\\
\langle v_4| \equiv &\frac{1}{\sqrt{8}} (\sqrt{2},0,0, -\sqrt{2},-1,\sqrt{3},0,0) ,\\
\langle v_5| \equiv &\frac{1}{\sqrt{8}} (\sqrt{2}, 0, -\sqrt{2}, 0, -1, -\sqrt{3},0,0),\\
\langle v_6| \equiv &\frac{1}{\sqrt{8}} (\sqrt{2}, 0, \sqrt{2}, 0, -1,-\sqrt{3},0,0), \\
\langle v_7| \equiv &\frac{1}{2} (1, 1, 1, -1,0,0,0,0),\\
\langle v_8| \equiv &\frac{1}{\sqrt{8}} (\sqrt{2}, \sqrt{2}, 0, 0, 2,0,0,0),\\
\langle v_9| \equiv &\frac{1}{2} (1, 1, -1, 1,0,0,0,0).
\end{align}
\label{eq:vis}
\end{subequations}
The state $|\psi\rangle$ on which the measurements 
$\mathcal{M}_i$ will be performed is chosen as
\begin{equation}
\langle\psi| \equiv (1,0,0,0,0,0,0,0),
\label{eq:state}
\end{equation}
so that $\langle v_i|\psi\rangle = \frac{1}{2}$ $\forall \, i\in\lbrace
0,1,...,9\rbrace$ which subsequently ensures the exclusivity relation $P(\Pi_i
+\Pi_{i+1} +\Pi_{i+5} +\Pi_{i+7}=1)=1$, $i=0,1,...,4$.

In order to evaluate the KCBS-twin inequality experimentally, we first
decompose the projectors involved in terms of Pauli operators, $\sigma_j$,
$j\in \lbrace x,y,z\rbrace$ for three qubits given by:
\begin{widetext}
\begin{subequations}
\begin{align}
\Pi_0 &= \frac{1}{16}\left(
\begin{aligned}
&-A_{0}+A_{1}+2A_{6} - A_{7}+A_{8}+\sqrt{2}(A_{9}
-A_{10}+A_{11}+A_{17}-A_{18}+ A_{19}-A_{20}-A_{25})\\
&\quad-A_{27}-A_{28}-A_{34}-A_{35}+2 \mathds{1}
\end{aligned}\right),\\
\Pi_1 &= \frac{1}{32}\left(
\begin{aligned}
&-\sqrt{3}A_{0}-A_{1}+2A_{3}-2A_{5}+2A_{6}-\sqrt{3}A_{7}
+A_{8}\\
&\quad+
\sqrt{2}\left(-A_{9}+\sqrt{6}A_{10}-A_{11}+\sqrt{6}A_{12}
-A_{13} -\sqrt{6}A_{14}+ A_{15}+ A_{16}-A_{17}+\sqrt{6} A_{18}\right)
\\
&\quad+\sqrt{2}\left(-A_{19}-\sqrt{6}A_{20}-A_{21}+\sqrt{6}A_{22}-A_{23}-\sqrt{6} A_{24}
-\sqrt{6}A_{25}\right)
\\ &\quad+\sqrt{3}A_{27}+A_{28}+2A_{30}-2A_{32}
-2A_{33}+\sqrt{3}A_{34}+3A_{35}+4\mathds{1}
\end{aligned}\right),\\
\Pi_2 &=  \frac{1}{8}\left(\begin{aligned}
-A_{4}+A_{5}-A_{7}+A_{26}-A_{31}+A_{32}-A_{34}+\mathds{1}
\end{aligned}\right),\\
\Pi_3 &=  \frac{1}{8}\left(A_{4}-A_{5}-A_{7}+A_{26}+A_{31}-A_{32} -A_{34}+\mathds{1}\right),\\
\Pi_4 &=  \frac{1}{32}
\left(\begin{aligned}
&-\sqrt{3}A_{0}-A_{1}-2A_{3}+2A_{5}+2A_{6}-\sqrt{3}A_{7}+A_{8}\\
&\quad+\sqrt{2}\left(-A_{9}+\sqrt{6}A_{10}-A_{11}-\sqrt{6}A_{12}
+A_{13} +\sqrt{6}A_{14}- A_{15}- A_{16}-A_{17}+\sqrt{6}
A_{18}\right)\\
&\quad+\sqrt{2}\left(-A_{19}-\sqrt{6}A_{20}+A_{21}-\sqrt{6}A_{22}+A_{23}+\sqrt{6}
A_{24}
-\sqrt{6}A_{25}\right)\\
&\quad+\sqrt{3}A_{27}+A_{28}-2A_{30}+2A_{32}
-2A_{33}+\sqrt{3}A_{34}+3A_{35}+4\mathds{1}\end{aligned}\right),\\
\Pi_5 &=  \frac{1}{32}\left(\begin{aligned}
&\sqrt{3}A_{0}+A_{1}-2A_{2}-2A_{4}+2A_{6}+\sqrt{3}A_{7} -A_{8}\\
&\quad+\sqrt{2}\left(-A_{9}-\sqrt{6}A_{10}-A_{11}+A_{12}
+\sqrt{6}A_{13} +A_{14}+ \sqrt{6}A_{16}-A_{17}-\sqrt{6}
A_{18}\right)\\
&\quad+\sqrt{2}\left(\-A_{19}+\sqrt{6}A_{20}-\sqrt{6}A_{21}+A_{22}+\sqrt{6}A_{23}+
A_{24}+\sqrt{6}A_{25}\right)\\
&\quad-\sqrt{3}A_{27}+3A_{28}-2A_{29}-2A_{31}-2A_{33}-\sqrt{3}A_{34}+A_{35}+4\mathds{1}
\end{aligned}\right),\\
\Pi_6 &=  \frac{1}{32}\left(\begin{aligned}
&\sqrt{3}A_{0}+A_{1}+2A_{2}+2A_{4}+2A_{6}+\sqrt{3}A_{7}-A_{8}\\
&\quad+\sqrt{2}
\left(-A_{9}-\sqrt{6}A_{10}-A_{11}-A_{12}
-\sqrt{6}A_{13} -A_{14}- \sqrt{6}A_{16}-A_{17}-\sqrt{6}
A_{18}\right)\\
&\quad+\sqrt{2}\left(
-A_{19}+\sqrt{6}A_{20}+\sqrt{6}A_{21}-A_{22}-\sqrt{6}A_{23}- 
A_{24}+\sqrt{6}A_{25}\right)\\
&\quad-\sqrt{3}A_{27}+3A_{28}+2A_{29}+2A_{31}
-2A_{33}-\sqrt{3}A_{34}+A_{35}+4\mathds{1}
\end{aligned}\right),\\
\Pi_7 &=  \frac{1}{8}\left(A_{4}+A_{5}+A_{7}+A_{26}+A_{31}+A_{32}+A_{34}+\mathds{1}\right),\\
\Pi_8 &=  \frac{1}{16}\left(\begin{aligned}
&+A_{0}+A_{1}+2A_{6} + A_{7}+A_{8}\\
&\quad+\sqrt{2}\left(A_{9} +A_{10}+A_{11}+A_{17}+A_{18}+ A_{19}+A_{20}+A_{25}\right)\\
&\quad  +A_{27}-A_{28}+A_{34}-A_{35}+2\mathds{1}
\end{aligned}\right),\\
\Pi_9 &=  \frac{1}{8}\left(-A_{4}-A_{5}+A_{7}+A_{26}-A_{31}-A_{32}+A_{34}+\mathds{1}\right).
\end{align}
\label{eq:proj_decomp}
\end{subequations}
\end{widetext}
where $A_i$~s are given in Table~\ref{tab:ais}.
In NMR, the observed $z$
magnetization of a nuclear spin in a quantum state is proportional to the
expectation value of the $\sigma_z$  operator of the spin in that state. The
time-domain NMR signal, i.e., the free-induction decay with appropriate phase
gives Lorentzian peaks when Fourier transformed. These normalized experimental
intensities give an estimate of the expectation value of $\sigma_z$ of the
quantum state. 
The observables of interest are 
$A_1, A_6, A_8, A_{26}, A_{28},
A_{33}, A_{35} $ for the eight-dimensional Hilbert space being considered.
The task of experimentally demonstrating the inequality
$\mathcal{K}$ (given in Eqn.~\ref{eq:kcbs_twin}) 
on an NMR quantum information processor
becomes particularly accessible while dealing with the Pauli basis, 
since the NMR
signal is proportional to ensemble average of the $\sigma_z$ operator. Thus
measurement of the expectation value of the projectors $\lbrace \Pi_i\rbrace$
involved becomes simplified when they are decomposed into Pauli
operators~\cite{singh-pra-18,gaikwad-pra-18,dogra-pla-16} given by the
observables $\lbrace A_i\rbrace$.

\begin{figure}[h]
\centering
\includegraphics[scale=1.0]{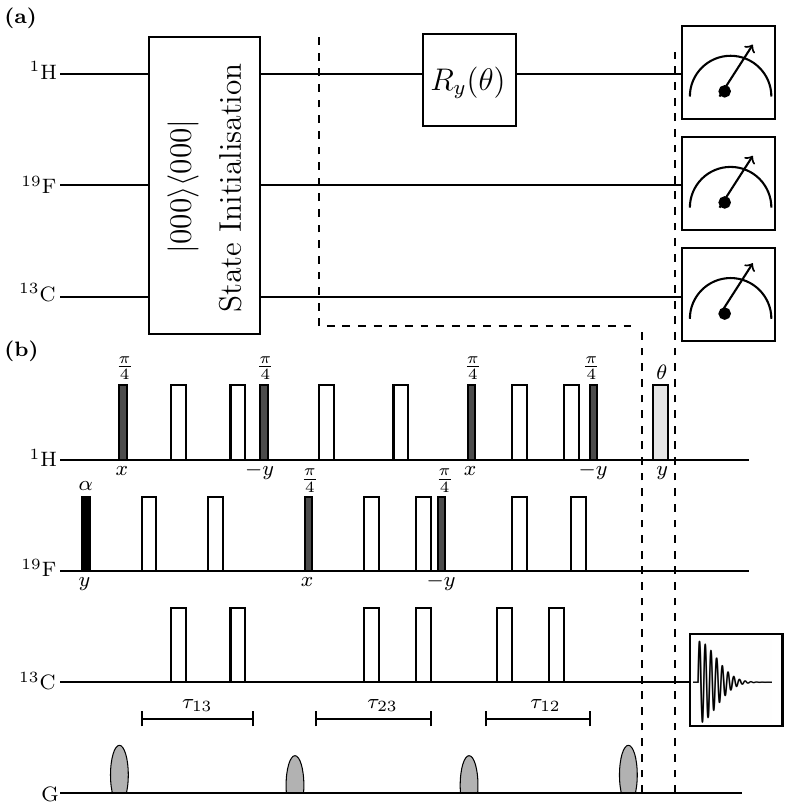}
\caption{(a) Quantum circuit for state preparation; the parameter
$\theta$ in the unitary
$R_y(\theta)$ is used to generate different quantum states.
(b) Corresponding NMR pulse sequence for the quantum circuit. The sequence of
pulses before the first dashed black line achieves initialization 
of the state into the pseudopure
$\vert 000 \rangle$ state. 
The value of the flip angle $\alpha$ is kept fixed at
$57.87^{\circ}$, while the
flip angle $\theta$ is varied over a range of values. The
broad unfilled rectangles denote $\pi$ pulses, and the flip angle 
and phases of the other pulses
written below each pulse. The
time intervals $\tau_{12}$, $\tau_{13}$, $\tau_{23}$ are set to 
$\frac{1}{2J_{HF}}$, $\frac{1}{2J_{HC}}$, $\frac{1}{2J_{FC}}$, respectively.}
\label{ckt-3qubit}
\end{figure}

Using the decomposition given in Eqn.~(\ref{eq:proj_decomp}), the inequality
$\mathcal{K}$ (given in Eqn.~\ref{eq:kcbs_twin}) can be re-written as:
\begin{equation}
\mathcal{K}=\frac{1}{8}\text{Tr}\left[A\cdot\rho\right]\overset{\text{NCHV}}{\leq}2
\overset{\text{QM, GP}}{\leq}\frac{5}{2},
\label{eq:kcbs_twin2}
\end{equation}
where $\rho=|\psi\rangle\langle\psi|$ and 
\begin{equation}
A=A_1+ 4A_6+ A_8+ 4A_{26}+ A_{28}-2A_{33}+ A_{35} + 10\mathds{1}.
\label{eq:form_a}
\end{equation}
By experimentally measuring the expectation value of each observable $A_i$ for
the state $\rho$, the value of the inequality $\mathcal{K}$ can be estimated.
The explicit mapping of expectation value of the observables onto
Pauli $z$ operators for three qubits is given in Table~\ref{3qubit-Pauli-table}.
The underlying state $\vert\psi\rangle$ is
unitarily rotated by an angle $\theta$ as:
\begin{equation}
\vert\psi(\theta)\rangle =
U_\theta\otimes\mathds{1}\otimes\mathds{1}\vert\psi\rangle,
\end{equation}
where,
\begin{equation}
U_\theta = \begin{bmatrix}
\cos\frac{\theta}{2} & -\sin \frac{\theta}{2}\\
\sin \frac{\theta}{2} & \cos \frac{\theta}{2}
\end{bmatrix}
\label{eq:rotation}
\end{equation}

\begin{figure}[h]
\centering
\includegraphics[scale=1.0]{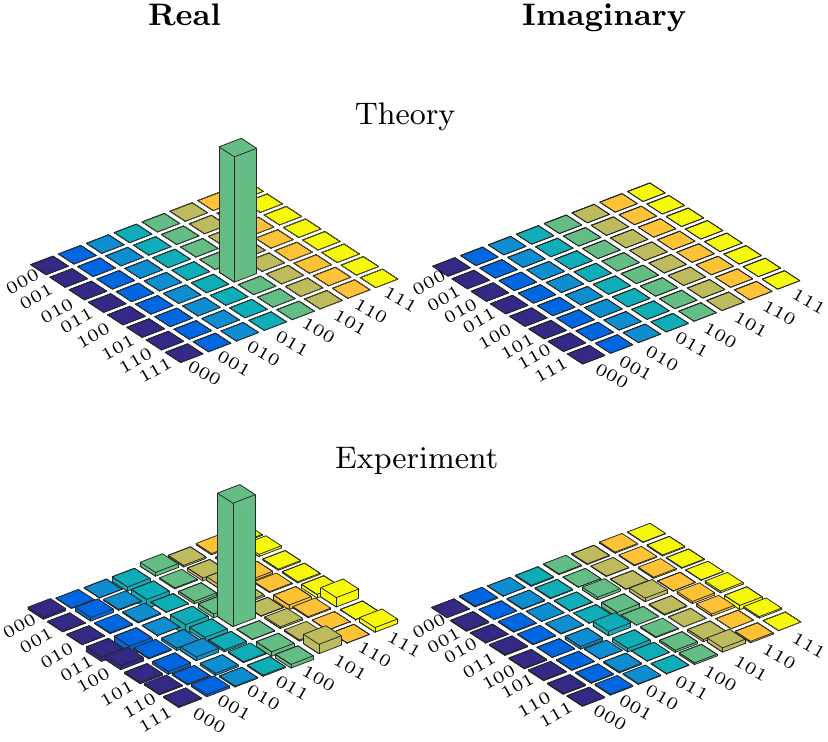}
\caption{Real (left) and imaginary (right) parts of the theoretically
expected and the 
experimentally reconstructed 
tomographs of the  $\langle \psi_1 \vert = (0,0,0,0,1,0,0,0) $
state in the eight-dimensional quantum system, with 
an  experimental state fidelity of
0.97.}
\label{tomo8dim}
\end{figure}

To experimentally implement the KCBS-twin inequality capable of revealing fully contextual
quantum correlations for an eight-dimensional quantum system,
we used the molecule of 
${}^{13}$C -labeled diethyl fluoromalonate dissolved in
acetone-D6, with the ${}^{1}$H, ${}^{19}$F and ${}^{13}$C spin-1/2
nuclei being encoded as `qubit one', `qubit two' and `qubit three', 
respectively (see Fig~\ref{molfig-3qubit} for the molecular
structure and corresponding NMR spectrum of the PPS state, and 
Table~\ref{parameter-table-3qubit}
for details of the experimental NMR parameters).  The
NMR Hamiltonian for a three-qubit system is given by~\cite{singh-pra-18}
\begin{equation}
\mathcal{H}=-\sum_{i=1}^{3} v_i I^i_z + \sum_{i>j,i=1}^{3} J_{ij} I^i_z I^j_z 
\end{equation}
where the indices $i, j$ = 1, 2, or 3 label the qubit, 
$\nu_{i}$ is the chemical shift of the
$i$th qubit in the rotating frame, $J_{{ij}}$ is the
scalar coupling interaction strength, and $I_z^{{i}}$ is 
$z$-component of the spin angular
momentum operator of the $i^{th}$ qubit. The system was initialized in a
pseudopure state (PPS), i.e., $ \vert 000 \rangle$, using the spatial
averaging technique~\cite{mitra-jmr-97} with the density operator given by
\begin{equation}
\rho_{000}=\frac{1-\epsilon}{2^3} \mathds{I}_8 + \epsilon  \vert 000 \rangle \langle 000 \vert 
\end{equation}
where $\epsilon$ is proportional to the spin polarization and $\mathds{I}_8$ is
the $8\times8$ identity operator. The fidelity of the experimentally prepared
PPS state was computed to be 0.96 using the fidelity 
measure~\cite{zhang-prl-14}.  Full quantum state tomography (QST)~\cite{leskowitz-pra-04,singh-pla-16} was performed
to experimentally reconstruct the density operator via a set of preparatory
pulses  $\{III, IIY, IYY, YII, XYX, XXY, XXX\}$, where $I$ implies no
operation, and $X(Y)$ denotes a qubit-selective rf pulse of flip angle
$90^{\circ}$ of phase $x(y)$. 

Experiments were performed at room temperature (294 K) on a Bruker Avance III
600-MHz FT-NMR spectrometer equipped with a QXI probe.  Local unitary
operations were achieved by using highly accurate and calibrated spin selective
transverse rf pulses of suitable amplitude, phase, and duration.  Nonlocal
unitary operations were achieved by free evolution under the system
Hamiltonian, of suitable duration under the desired scalar coupling with the
help of embedded $\pi$ refocusing pulses.  The durations of the $\frac{\pi}{2}$
pulses for ${}^{1}$H, ${}^{19}$F, and ${}^{13}$C nuclei were 9.36 $\mu$s at
18.14 W power level, 23.25 $\mu$s at a power level of 42.27 W, and 15.81 $\mu$s
at a power level of 179.47 W, respectively.

\begin{figure}[h]
\centering
\includegraphics[scale=1.0]{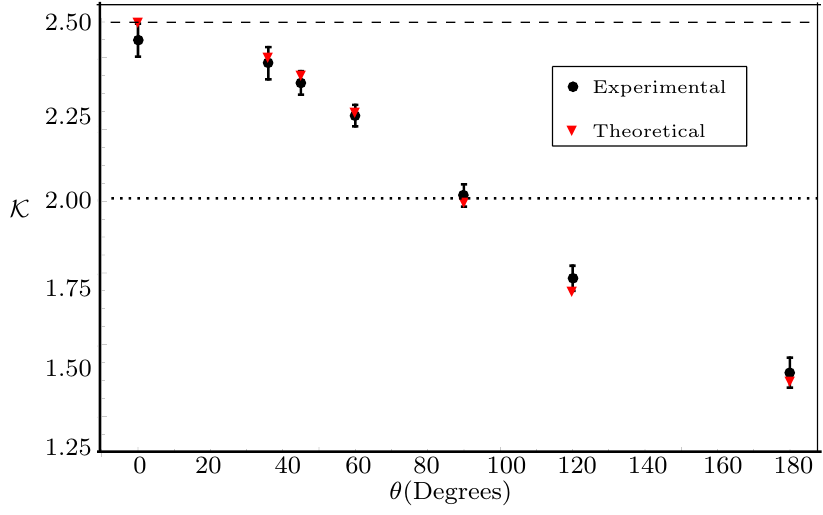}
\caption{Graph representing quantum correlations corresponding to the
inequality $\mathcal{K}$ for various states rotated by angle $\theta$ from the
initial state $|\psi\rangle$.} 
\label{plot-3qubit}
\end{figure}
\begin{table}[h]
\caption{Theoretically computed and experimentally 
measured values of quantum correlations 
corresponding to the inequality  $\mathcal{K}$ for various states, 
rotated by angle $\theta$, from the initial state $|\psi\rangle$.}  
\vspace*{12pt}
\centering
\begin{tabular}{ccc}
\hline
$\theta$ & Theoretical  & 
Experimental\\
\hline
180$^{\circ}$ & 1.500 & 1.522$\pm$0.042 \\
120$^{\circ}$ & 1.750 & 1.785$\pm$0.035 \\
90$^{\circ}$ & 2.000 & 2.016$\pm$0.031 \\
60$^{\circ}$ & 2.250 & 2.239$\pm$0.030 \\
45$^{\circ}$ & 2.353 & 2.330$\pm$0.033  \\
36$^{\circ}$ & 2.404 & 2.385$\pm$0.045  \\
0$^{\circ}$ & 2.500 &  2.449$\pm$0.046 \\
\hline
\end{tabular}
\label{3qubit-value-table}
\end{table}

The quantum circuit to construct the states required to test fully contextual
quantum correlations is shown in Fig.~\ref{ckt-3qubit}(a) 
and the corresponding NMR pulse
sequence is shown in Fig~\ref{ckt-3qubit}(b).  
Different states can be prepared by varying
the value of the flip angle $\theta$ of the rf pulse.  
We prepared seven different
states by varying the flip angle $\theta$ to attain a range of
values: $180^{\circ}, 120^{\circ}, 90^{\circ}, 60^{\circ}, 
45^{\circ}, 36^{\circ}, 0^{\circ}$. The state prepared with
$\theta=180^{\circ}$  gives the minimum value of  $\mathcal{K}$,
while the
state prepared without applying any rf pulse ($\theta=0^{\circ}$) gives the
maximum value.
All the states required to demonstrate the KCBS-twin inequality 
on an 8-dimensional Hilbert space which are
capable of revealing the transformation from classical correlations to fully
contextual correlations, were experimentally  prepared with state fidelities of
$\geq 0.96$. 
The tomograph of one such experimentally reconstructed 
state with 
flip angle $\theta =180^{\circ}$ with 
state fidelity 0.97 is depicted in Fig.~\ref{tomo8dim}. 
For each of the initial states, the
contextuality test was repeated three times. 
The mean values and the corresponding
error bars were computed and the result is shown in Fig~\ref{plot-3qubit},
where the inequality values are plotted for different values of the parameter
$\theta$.  The maximum of the sum of probabilities using classical theory is
$2$ and the maximum of sum of probabilities using quantum theory is $2.5$,
which are depicted by dotted and dashed lines respectively in
Fig~\ref{plot-3qubit}. The theoretically computed and experimentally obtained
values of the inequality for different values of the $\theta$ parameter are
tabulated in Table~\ref{3qubit-value-table}.  The theoretical and experimental
values match well, within the limits of experimental errors.  From
Fig.~\ref{plot-3qubit} it is also seen that the violation observed for the
KCBS-twin inequality decreases as the original state $|\psi\rangle$ is rotated
through an angle $\theta$, with no violation when the transformed state is
orthogonal to the original state. Furthermore, the plot is nonlinear,
indicating that smaller rotations lead to minor changes in violation, while
larger rotations may also lead to observing no violation at all.
\section{Fully contextual quantum correlations in a 
four-dimensional Hilbert space}
\label{ineq2}   
\begin{figure}[h]
\centering
\includegraphics[scale=1.0]{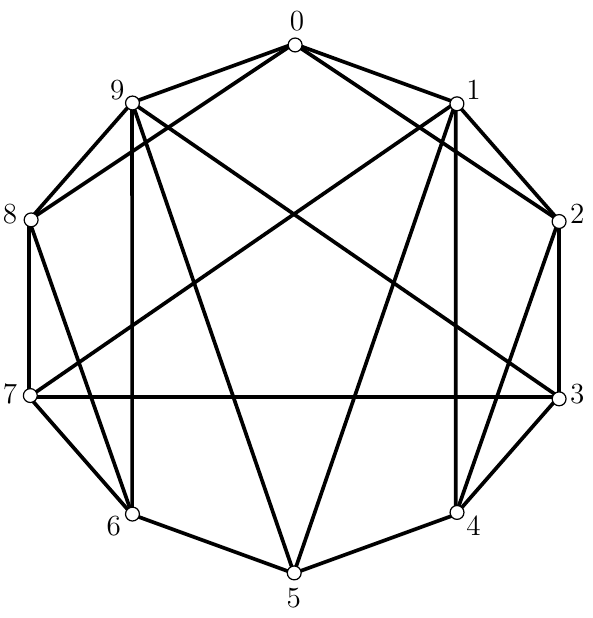}
\caption{Orthogonality graph corresponding to the inequality $\mathcal{C}$.
Vertices correspond to projectors and two vertices are connected by an edge if
they are orthogonal.}
\label{fig-fullcontext}
\end{figure}
In this section, we first review a contextuality
inequality which is capable of revealing 
fully contextual quantum correlations as
developed by 
Nagali {\em et. al}~\cite{nagali-prl-12} which
utilizes states in a Hilbert space of dimension at least four. 
We provide a
modified version of the inequality by decomposition into Pauli matrices
which we experimentally test on a four-level quantum system
using two NMR qubits.  
Fully contextual quantum correlations can also be achieved for
scenarios other than KCBS. As shown in Reference~\cite{nagali-prl-12}, one
such scenario entails measurements corresponding to ten different projectors
$\Pi_j=|u_j\rangle\langle u_j|$, $j=\lbrace0,1,...,9\rbrace$.  In this
particular scenario, the projectors follow exclusivity relationships as
depicted in Fig.~\ref{fig-fullcontext}, where each vertex represents a
projector $\Pi_i$ and two projectors are connected by an edge if and only if
they are exclusive. The corresponding test of contextuality is then given by
the inequality: 
\begin{equation} 
\mathcal{C}=\sum_{i=0}^{9}
P(\Pi_i=1)\overset{\text{NCHV}}{\leq} 3\overset{\text{QM,
GP}}{\leq}\frac{7}{2}.  
\label{eq:context_test} 
\end{equation}

\begin{table}[h]
\caption{Product operators for a two-qubit
system mapped onto the Pauli $z$ operators via the
initial state $\rho \rightarrow \rho_i=U_i.\rho.U^\dagger_i$.}
\vspace*{12pt}
\centering
\begin{tabular}{rp{12pt}l}
\hline
Observable Expectation  &&
Unitary Operator\\
\hline
 $\langle \sigma_{1x}\sigma_{2x} \rangle$ = Tr[$\rho_{1}.
\sigma_{2z}$] && $U_1={\rm{CNOT_{12}~  }}Y_2Y_1$ ~~~\\
  $\langle \sigma_{1y}\sigma_{2y} \rangle$ =
Tr[$\rho_{2}.\sigma_{2z}$] & &$U_{2}={\rm{CNOT_{12}~  }}\overline{X}_2\overline{X}_1$ ~~~\\
 $\langle \sigma_{1z} \rangle$ = Tr[$\rho_{3}.\sigma_{1z}$].
&& $U_{3}$=Identity~~~\\
  $\langle \sigma_{1z}\sigma_{2z} \rangle$ = 
Tr[$\rho_{4}.\sigma_{2z}$]&&  $U_{4}=\rm{CNOT_{12}~  }$  ~~~\\
 $\langle \sigma_{2z} \rangle$ = Tr[$\rho_{5}.\sigma_{2z}$]
&&  $U_{5}$=Identity ~~~\\
\hline
\end{tabular}
\label{2qubit-Pauli-table}
\end{table}

The scenario is reminiscent of the KCBS-twin inequality discussed
in the previous section,
however this test requires ten different measurements rather than five and is
capable of revealing fully contextual quantum correlations in a much smaller
Hilbert space (of minimum dimension four). 
The inequality can be explicitly
tested if we consider the unit vectors $|u_i\rangle$ as follows:
\begin{subequations}
\begin{align}
\langle u_0| \equiv &\frac{1}{\sqrt{2}}(0,0,1,1), \\
\langle u_1| \equiv &\frac{1}{2}(1,-1,1,-1), \\
\langle u_2| \equiv &\frac{1}{2}(1,-1,-1,1),\\
\langle u_3| \equiv &\frac{1}{\sqrt{2}}(1,0,0,-1),\\
\langle u_4| \equiv &\frac{1}{2}(1,1,1,1) ,\\
\langle u_5| \equiv &\frac{1}{\sqrt{2}}(0,1,0,-1),\\
\langle u_6| \equiv &\frac{1}{2}(-1,1,1,1) ,\\
\langle u_7| \equiv & \frac{1}{\sqrt{2}}(1,0,0,1),\\
\langle u_8| \equiv &\frac{1}{2}(1,1,1,-1) ,\\
\langle u_9| \equiv &\frac{1}{2}(1,1,-1,1).
\end{align}
\label{eq:uis}
\end{subequations}
The corresponding projective measurements are of the form
\begin{equation}
\mathcal{M}_j=\lbrace\Pi_j,\mathds{1}-\Pi_j\rbrace \quad \forall \, j\in\lbrace0,1,...,9\rbrace,
\label{eq:meas}
\end{equation}
which are performed on the state
\begin{equation}
\langle \phi \vert \equiv (0,0,0,1).
\label{eq:state2}
\end{equation}

For the experimental implementation of the inequality, we again 
decompose the 
projectors $\lbrace\Pi_j\rbrace$ in terms of Pauli 
operators :
\begin{subequations}
\begin{align}
\Pi_0=&\frac{1}{4}(-\sigma_z\otimes \mathds{1}- \sigma_z\otimes \sigma_x+ 
\mathds{1} \otimes \sigma_x+ \mathds{1}\otimes\mathds{1}), \\
\Pi_1=&\frac{1}{4}(\sigma_x\otimes \mathds{1}- \sigma_x\otimes \sigma_x- 
\mathds{1} \otimes \sigma_x+ \mathds{1}\otimes\mathds{1}), \\
\Pi_2=&\frac{1}{4}(-\sigma_x\otimes \mathds{1}+ \sigma_x\otimes \sigma_x- 
\mathds{1} \otimes \sigma_x+ \mathds{1}\otimes\mathds{1}),\\
\Pi_3=&\frac{1}{4}(-\sigma_x\otimes \sigma_x+ \sigma_y\otimes \sigma_y+ 
\sigma_z\otimes \sigma_z+ \mathds{1}\otimes\mathds{1}),\\
\Pi_4=&\frac{1}{4}(\sigma_x\otimes \mathds{1}+ \sigma_x\otimes \sigma_x+ 
\mathds{1} \otimes \sigma_x+ \mathds{1}\otimes\mathds{1}),\\
\Pi_5=&\frac{1}{4}(-\sigma_x\otimes \mathds{1}+ \sigma_x\otimes \sigma_z- 
\mathds{1} \otimes\sigma_z+ \mathds{1}\otimes\mathds{1}),\\
\Pi_6=&\frac{1}{4}(-\sigma_x\otimes \sigma_z+\sigma_y\otimes \sigma_y- 
\sigma_z\otimes \sigma_x+ \mathds{1}\otimes\mathds{1}),\\
\Pi_7=&\frac{1}{4}(\sigma_x\otimes \sigma_x- \sigma_y\otimes \sigma_y+ 
\sigma_z\otimes \sigma_z+ \mathds{1}\otimes\mathds{1}),\\
\Pi_8=&\frac{1}{4}(\sigma_x\otimes \sigma_z+ \sigma_y\otimes \sigma_y+ 
\sigma_z\otimes \sigma_x+ \mathds{1}\otimes\mathds{1}),\\
\Pi_9=&\frac{1}{4}(-\sigma_x\otimes \sigma_z- \sigma_y\otimes \sigma_y+ 
\sigma_z\otimes \sigma_x+ \mathds{1}\otimes\mathds{1}).
\end{align}
\label{eq:u_proj_sigma}
\end{subequations}

Using Eqn.~(\ref{eq:context_test}) and Eqn.~(\ref{eq:u_proj_sigma}), 
the inequality $\mathcal{C}$ can be re-written as 
\begin{equation}
\mathcal{C}=\frac{1}{4}\text{Tr}\left[B\cdot\rho'\right]\overset{\text{NCHV}}{\leq} 3\overset{\text{QM, GP}}{\leq}\frac{7}{2},
\label{eq:context_test2}
\end{equation}
where $\rho'=|\phi\rangle\langle\phi|$ and
\begin{equation}
B=B_0+ B_1-B_2+2B_3-B_4 +10\mathds{1},
\label{eq:form_b}
\end{equation}
with
\begin{equation}
\begin{aligned}
B_0&=\sigma_x\otimes\sigma_x,\\
B_1&=\sigma_y\otimes\sigma_y,\\
B_2&=\sigma_z\otimes \mathds{1},\\
B_3&=\sigma_z\otimes \sigma_z,\\
B_4&=\mathds{1} \otimes\sigma_z.
\end{aligned}
\end{equation}
The underlying state $\vert\phi\rangle$ is
unitarily rotated by an angle $\theta$ as:
\begin{equation}
\vert\phi(\theta)\rangle = U_\theta\otimes\mathds{1}\vert\phi\rangle,
\end{equation} 
where $U_\theta$ has been defined in Eqn.~(\ref{eq:rotation}).

\begin{figure}[h]
\centering
\includegraphics[scale=1.0]{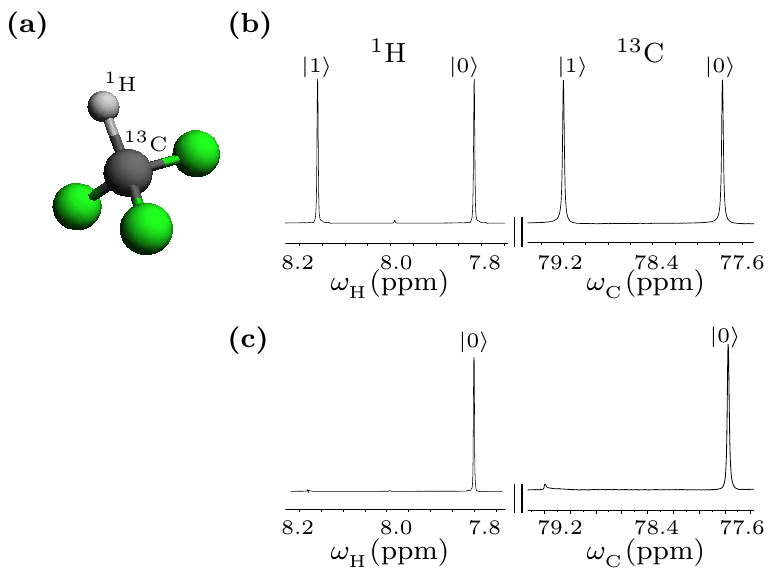}
\caption{(a) Molecular structure of ${}^{13}$C-labeled chloroform used as a
two-qubit quantum system. NMR spectra of (b) the thermal equilibrium state 
and (c) the
pseudopure state $\vert 00 \rangle$.
Each peak is labeled with the
logical state of the qubit which is passive during the transition. Horizontal
scale represents the chemical shifts in ppm.}
\label{molfig-2qubit}
\end{figure}

By experimentally evaluating the expectation value of the observables $B_j$,
the value of the inequality $\mathcal{C}$ can be estimated.
To implement the non-contextual inequality capable of revealing fully contextual
quantum correlations on a four-dimensional quantum 
system, the molecule of 
${}^{13}$C-enriched chloroform dissolved in acetone-D6 was used, with the
${}^{1}$H and ${}^{13}$C  spins being labeled as `qubit one' and `qubit
two', respectively (see Fig.~\ref{molfig-2qubit} and 
Table~\ref{2qubit-parameter-table} for details of the
experimental parameters).
\begin{table}[h]
\caption{NMR parameters for ${}^{13}$C-labeled
chloroform used as a two-qubit quantum system.}
\vspace*{12pt}
\centering
\renewcommand{\arraystretch}{1.7}
\begin{tabular}{lllll}
\hline
Qubit & 
~$\nu$ (Hz)& $~J$ (Hz) & $T_1$ (sec) & $T_2$ (sec)~~~\\
\hline
$\rm ^{1}H$ & $4787.86$ & $~J_{HC}=215.11$ & $~7.9$ & $2.95$~~~\\
$\rm ^{13}C$ & $11814.09$ & & $~16.6$ & $0.3$ ~~~\\
\hline
\end{tabular}
\label{2qubit-parameter-table}
\end{table}
The Hamiltonian for a two-qubit system is given by~\cite{gaikwad-pra-18} 
\begin{equation}
{\mathcal{H}}= -\nu_{{\rm H}} I_z^{{\rm H}} - 
\nu_{{\rm C}} I_z^{{\rm C}} + 
J_{{\rm HC}} I_z^{{\rm H}} I_z^{{\rm C}}
\end{equation}
where $\nu_{{\rm H}}$, $\nu_{{\rm C}}$ are the chemical shifts, $I_z^{{\rm
H}}$, $I_z^{{\rm C}}$ are the $z$-components of the spin angular momentum
operators of the ${}^{1}$H and ${}^{13}$C spins respectively, and J$_{{\rm HC}}$
is the scalar coupling constant. The system was initialized in the pseudopure
state (PPS)  $\vert 00 \rangle$, using the spatial averaging 
technique~\cite{oliveira-book-07,singh-pra-17} with the
density operator given by 
\begin{equation} 
\rho_{00}=\frac{1}{4}(1-\epsilon)\mathds{I}_4+\epsilon
\vert 00\rangle \langle 00 \vert 
\end{equation}
where  $\mathds{I}_4$ is the $4\times4$ identity operator, $\epsilon$ is
proportional to the spin polarization and 
can be evaluated from the ratio of magnetic
and thermal energies of an ensemble of magnetic moments $\mu$ in a magnetic
field $B$ at temperature $T$; $\epsilon \thicksim \frac{\mu B}{k_B T}$ and at
room temperature  and for a $B \approx$ 10 Tesla, $\epsilon \approx
\rm{10^{-5}}$. The state fidelity of the experimentally prepared PPS 
was computed to be 0.99. For the experimental reconstruction of density
operator full quantum state tomography (QST) was performed using a set of
preparatory pulses  $\{II, IX, IY, XX\}$.  
Most of the experimental details are the same as for the three-qubit case.
The durations of $\frac{\pi}{2}$ pulses for ${}^{1}$H, ${}^{13}$C 
nuclei were 9.56
$\mu$s at power level  18.14 W and 16.15 $\mu$s at a power level of 179.47 W,
respectively. 
\begin{figure}[h]
\centering
\includegraphics[scale=1.0]{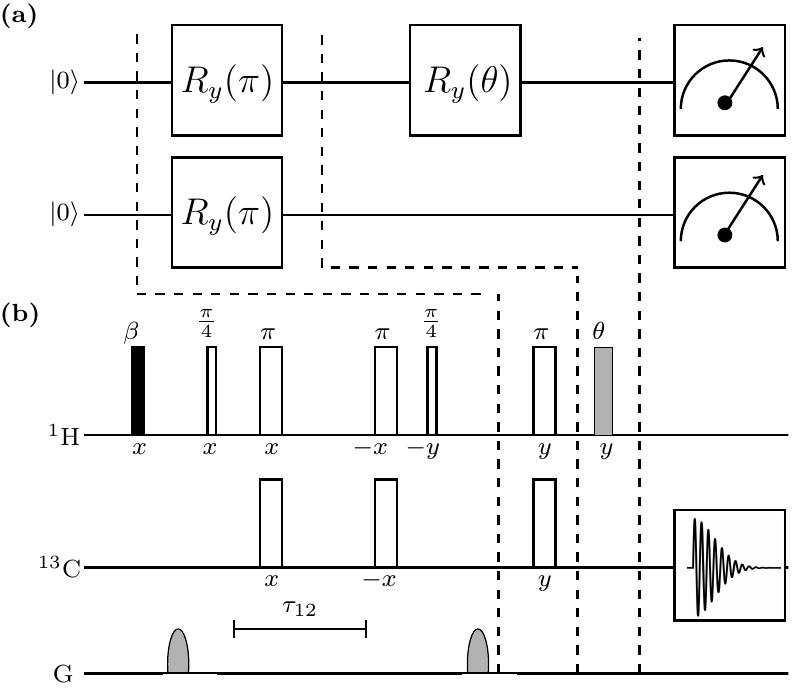}
\caption{(a) Quantum circuit for the required state, generated randomly
generated with the different flip angles.
(b) NMR pulse sequence for the corresponding quantum circuit. The sequence of
pulses before the first dashed black line achieves state initialization into the
$\vert 00 \rangle$ state. The value of the flip angle $\beta$ 
is kept fixed at $59.69^{\circ}$, while the pulse rf
flip angle  $\theta$ is varied. The
time interval $\tau_{12}$ is set to $\frac{1}{2J_{HC}}$.}
\label{ckt-2qubit}
\end{figure}

Let $\pi_i$ be the observables (projectors) whose expectation value is to be
measured in a state $ \rho=\vert\psi\rangle\langle\psi\vert$. Instead of
measuring $\langle\pi\rangle$, the state $\rho$ can be mapped to $\rho_i$ by
using $\rho_i=U_i.\rho.U_i^\dagger$ followed by a $z$-magnetization measurement
of one of the qubits~\cite{gaikwad-pra-18}. 
Table~\ref{2qubit-Pauli-table} details the mapping of Pauli basis
operators (used in this paper) to the single-qubit Pauli $z$ operator, 
where $X, \overline{X}, Y$ and $\overline{Y}$ represent 
the $\frac{\pi}{2}$ rotations with phases $x, -x, y$ and $-y$, respectively. The observables of interest are $B_0, B_1, B_2, B_3, B_4$
for the four-dimensional Hilbert space under consideration. 
\begin{figure}[h]
\centering
\includegraphics[scale=1.0]{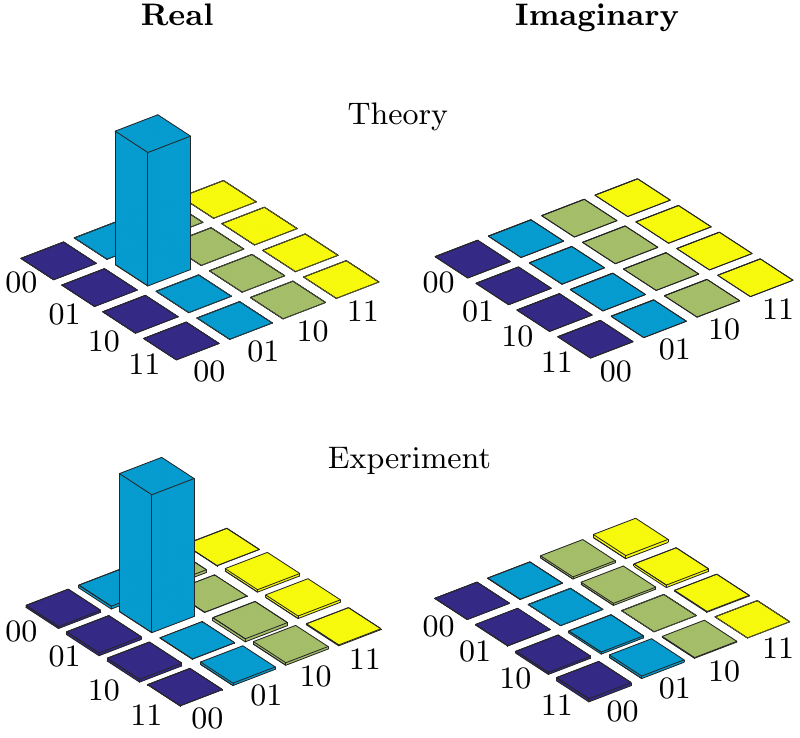}
\caption{Real (left) and imaginary (right) parts of the theoretical and
experimental tomographs of the  $\langle \phi_1 \vert = (0,-1,0,0) $  state 
in the four-dimensional Hilbert space, prepared with an
experimental state fidelity of 0.99.}
\label{tomo4dim}
\end{figure}

The quantum circuit to achieve the required states to test the inequality
$\mathcal{C}$ on a four-dimensional quantum system is shown in
Fig.~\ref{ckt-2qubit}(a) and the corresponding NMR pulse sequence is shown in
Fig.~\ref{ckt-2qubit}(b).  Eight different states were generated  by varying
the flip angle $\theta$ over a range of values: 
$180^{\circ}, 120^{\circ}, 90^{\circ}, 
69.23^{\circ}, 60^{\circ}, 45^{\circ},
30^{\circ}, 
0^{\circ}$. The state that is prepared with the 
flip angle $\theta=180^{\circ}$  gives the
minimum value of $\mathcal{C}$, while the state 
which is prepared without applying any rf pulse
($\theta=0^{\circ}$) gives the maximum value.
All the states required for testing the inequality on the four-dimensional 
quantum
system were experimentally  prepared with state fidelities $\geq 0.97$. The
tomograph for one such experimentally prepared state with the flip angle
$\theta=180^{\circ}$ and state fidelity 0.99 is depicted in
Fig.~\ref{tomo4dim}. 
\begin{figure}[h]
\centering
\includegraphics[scale=1.0]{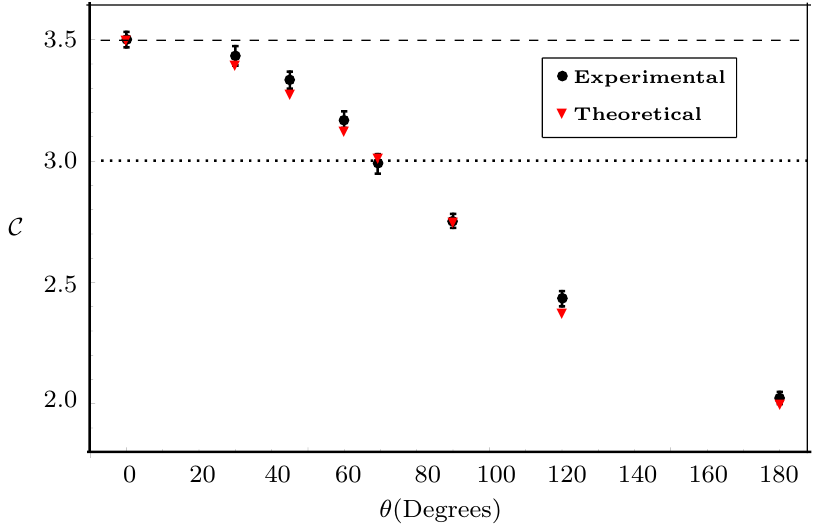}
\caption{Quantum correlations corresponding to the
inequality $\mathcal{C}$ for various states plotted for different
initial states $|\phi\rangle$, as a function of the
$\theta$ parameter.} 
\label{plot-2qubit}
\end{figure}
\begin{table}[h]
\caption{Theoretically computed  and experimentally measured 
values of quantum correlations corresponding to the 
inequality $\mathcal{C}$ for different quantum states 
parameterized by the 
angle $\theta$.}  
\vspace*{12pt}
\centering
\begin{tabular}{ccc}
\hline
$\theta$ & Theoretical  & 
Experimental\\
\hline
180$^{\circ}$ & 2.000 & 2.024$\pm$0.025 \\
120$^{\circ}$ & 2.375 & 2.433$\pm$0.031 \\
90$^{\circ}$ & 2.750 & 2.754$\pm$0.029 \\
69.23$^{\circ}$ & 3.016 & 2.989$\pm$0.040 \\
60$^{\circ}$ & 3.125 & 3.171$\pm$0.034 \\
45$^{\circ}$ & 3.280 & 3.334$\pm$0.035  \\
30$^{\circ}$ & 3.399 & 3.434$\pm$0.040  \\
0$^{\circ}$ & 3.500 &  3.501$\pm$0.032 \\
\hline
\end{tabular}
\label{2qubit-value-table}
\end{table}

For each of these eight different initial states, the contextuality
test was repeated three times. 
The mean values and the corresponding error bars were
calculated and result is shown in Fig.~\ref{plot-2qubit}, 
where the inequality values 
are plotted for different $\theta$ values. 
The maximum of sum of probabilities using
classical theory is $3$ and the maximum of sum of 
probabilities using quantum theory
is $3.5$, which are shown by dotted and dashed lines respectively in 
Fig~\ref{plot-2qubit}. As seen from the values tabulated in 
Table~\ref{2qubit-value-table}, the theoretically computed and
experimentally measured values of the inequality agree well to within
experimental errors.
From Fig.~\ref{plot-2qubit} it is seen that the violation for the
inequality $\mathcal{C}$ decreases as the original state $|\phi\rangle$ is
rotated through an angle $\theta$. It is seen that no violation is observed for
angle $\theta>70^\circ$, which is in contrast with the inequality
$\mathcal{K}$, which exhibits violation for a larger range of $\theta$.
However, certain similarities remain, most notably the nonlinear nature of
violation with respect to rotation.
It is
again observed that smaller rotations lead to minor changes in the
violation, while larger rotations may lead to a situation where no
violation is observed.
\section{Concluding Remarks}
\label{concl}
In this paper we experimentally demonstrated fully contextual quantum
correlations on an NMR quantum information processor. We studied two distinct
inequalities capable of revealing such correlations: the first inequality used
five measurements on an eight-dimensional 
Hilbert space, while the second
inequality used ten measurements on a four-dimensional Hilbert space 
to reveal the contextuality of the state.
However, both the inequalities involved the same number of
projectors. For an experimental demonstration of each inequality, every
projector was decomposed in terms of the Pauli basis, and the corresponding
inequality recast in terms of Pauli operators, thereby reducing the need for
resource-intensive full state tomography. Both the inequalities 
$\mathcal{K}$ and $\mathcal{C}$ were
experimentally implemented with a fidelity of $\geq 0.96$ by measuring the
expectation values of only seven and five Pauli operators for the state which
maximizes the violation, respectively.

In addition to demonstration of fully contextual quantum correlations, we
analyzed the behavior of each inequality under rotation of the underlying
state, which unitarily transforms it to another pure
state. The experiments were repeated for various states rotated through an
angle $\theta$ and were in good agreement with theoretical results. It was seen
that both the inequalities follow a nonlinear trend, while the inequality
$\mathcal{K}$ offers a greater range of violation than the
inequality $\mathcal{C}$ with respect to the parameter $\theta$.

An experimental implementation of fully contextual quantum correlations is an
important step towards achieving information processing tasks, for which no
post-quantum theory can do better. While the inequality $\mathcal{C}$ has been
experimentally observed on optical systems, an experimental demonstration of
the inequality $\mathcal{K}$ is difficult owing to the high dimensionality of
the Hilbert space required. Our work asserts the fact that NMR is an optimal
test bed for such scenarios.

\begin{acknowledgments}
All the experiments were performed on a Bruker Avance-III 600 MHz FT-NMR
spectrometer at the NMR Research Facility of IISER Mohali. J.S. acknowledges
funding from University Grants Commission, India.  Arvind acknowledges funding
from Department of Science and Technology, New Delhi, India under Grant No.
EMR/2014/000297. K.D. acknowledges funding from Department of Science and
Technology, New Delhi, India under Grant No. EMR/2015/000556.
\end{acknowledgments}

\bibliographystyle{apsrev4-1}

%

\end{document}